\documentclass{article}
\PassOptionsToPackage{numbers, compress}{natbib}
\usepackage[main, final]{neurips_2026}
\usepackage[utf8]{inputenc} % allow utf-8 input
\usepackage[T1]{fontenc}    % use 8-bit T1 fonts
\usepackage{tabularx}
\usepackage{array}
\usepackage{hyperref}       % hyperlinks
\usepackage{url}            % simple URL typesetting
\usepackage{booktabs}       % professional-quality tables
\usepackage{amsfonts}       % blackboard math symbols
\usepackage{nicefrac}       % compact symbols for 1/2, etc.
\usepackage{microtype}      % microtypography
\usepackage{xcolor}
\usepackage{graphicx}
\usepackage{amsmath}
\usepackage{placeins}
\usepackage{caption}
\usepackage{float}
\title{BACE: Behavior-Adaptive Connectivity Estimation from Multi-Region Neural Recordings}

\author{%
  \textbf{Mehrnaz Asadi}\textsuperscript{1},
  \textbf{Sina Javadzadeh}\textsuperscript{1},
  \textbf{Rahil Soroushmojdehi}\textsuperscript{1} \\
  \textbf{S.~Alireza Seyyed Mousavi}\textsuperscript{1},
  \textbf{Terence D.~Sanger}\textsuperscript{1,2} \\
  \textsuperscript{1}University of California, Irvine \\
  \textsuperscript{2}Rady Children's Health Orange County \\
  \texttt{Mehrnaza@uci.edu}
}
\begin{document}

\maketitle

\begin{abstract}
Understanding how distributed brain regions coordinate during behavior requires models that are both predictive and interpretable. We introduce Behavior-Adaptive Connectivity Estimation (BACE), an end-to-end framework for learning behavior-conditioned directed effective connectivity from multi-region intracranial local field potentials (LFPs). BACE first encodes within-region dynamics with region-specific temporal encoders, then applies a learned adjacency matrix selected for each behavioral context, and finally forecasts future neural activity through a graph-conditioned autoregressive decoder. This framework yields explicit region-level connectivity matrices whose edges are tied to predictive dynamics rather than post-hoc explanation. On synthetic time series with known directed graphs, BACE recovers the ground-truth edge structure from forecasting alone. We then evaluate BACE on human deep-brain LFP recordings from three participants performing a cued reaching task. Across participants, BACE achieves strong neural forecasting while producing compact, behavior-specific directed graphs across task contexts. Reliability analyses further support the stability of the inferred connectivity patterns. Together, these results position BACE as a practical framework for estimating behavior-conditioned effective connectivity from high-dimensional intracranial recordings, enabling interpretable hypotheses about how deep-brain regions interact during behavior.
\end{abstract}
\section{Introduction} 
Connectivity in neural time series is often modeled as a single graph between populations, without explicitly capturing how it varies across behavioral context. Connectivity is commonly described as structural (anatomical pathways), functional (statistical dependencies), and effective (directed influence) \citep{friston2011,fornito2016fundamentals}. However, these are not equally suited for this setting: structural connectivity is fixed, while functional connectivity captures undirected relationships typically estimated over aggregated time periods. As a result, methods for estimating directed connectivity that varies across behavioral conditions remain limited \citep{mohammadi2024graph,finc2020dynamic}.

Graph-based models are a common approach for representing connectivity in neural data, where nodes correspond to brain regions and edges encode relationships between them. This framework enables quantitative analysis of network organization, including modularity, hubs, and task-dependent variation \citep{bassett2017network,cole2013multi}. In practice, these graphs are typically constructed using pairwise correlations or related statistical measures computed over extended time periods. Consequently, they are often undirected and reflect average statistical dependencies between regions, rather than directed connectivity. This limits their ability to capture predictive dependencies and to provide interpretable representations of connectivity across contexts.

These limitations are particularly pronounced in intracranial local field potential (LFP) recordings across multiple deep-brain regions \citep{buzsaki2012,einevoll2013,kasiri2025optimized,kasiri2025deep}. Multi-region LFP data introduce additional modeling challenges: (i) each anatomical region is represented by multiple micro-contacts that must be aggregated into a coherent regional representation; (ii) neural activity evolves across well-defined behavioral epochs; and (iii) spatial organization is irregular and non-Euclidean \citep{bronstein2021}. These characteristics make it necessary to estimate connectivity at the level of regions and behavioral contexts, rather than at individual channels or over aggregated time periods.

Machine learning approaches have been used to infer connectivity
and interregional interactions directly from neural time series
\citep{perich2026curbd,li2026facrnn}, including through
forecasting-based objectives that capture predictive dependencies
in temporal data \citep{AMAG}. Graph-based architectures further incorporate spatial structure, enabling data-driven estimation of connectivity. Across these approaches, several challenges remain: many operate at the channel level, producing high-dimensional graphs that are difficult to interpret; others assume static or task-agnostic connectivity; and many represent dependencies implicitly through latent variables or attention mechanisms without exposing explicit, directed connectivity. As a result, methods that learn compact, directed, and context-dependent connectivity at the region level remain limited, particularly in settings where behaviorally meaningful structure must be identified.

In this work, we adopt a predictive view of connectivity: we define effective connectivity as the directed structure that improves prediction of future neural activity under a given behavioral context. This formulation links connectivity estimation directly to temporal prediction, rather than post-hoc analysis of correlations. Building on this principle, we introduce \textbf{Behavior-Adaptive Connectivity Estimation (BACE)}, an end-to-end framework that learns context-specific, directed inter-regional connectivity from multi-region intracranial LFP recordings. BACE  aggregates multiple micro-contacts within each anatomical region using independent temporal encoders to produce region-level representations, and learns a separate adjacency matrix for each behavioral context. These adjacency matrices are coupled to neural dynamics through a forecasting objective, ensuring that the learned connectivity reflects predictive dependencies.

We evaluate BACE on both synthetic and real datasets. On synthetic multivariate time series with known ground-truth graphs, BACE accurately recovers directed connectivity while maintaining competitive forecasting performance. Applied to human subcortical recordings from N regions during a cued reaching task, BACE produces explicit $N{\times}N$ connectivity matrices for each behavioral context, enabling direct comparison of connectivity patterns across conditions. Beyond predictive accuracy, we assess reproducibility and statistical reliability, treating learned adjacency matrices as explicit model outputs rather than latent representations.

Our contributions are summarized as follows:
\begin{itemize}
    \item \textbf{Context-dependent connectivity estimation.} We introduce a framework that explicitly models how directed connectivity varies across behavioral contexts, rather than assuming a fixed or task-agnostic graph.
    
    \item \textbf{Forecasting-based structure identification.} We show that predictive objectives can recover meaningful directed connectivity from neural data, aligning connectivity estimation with temporal prediction.
    
    \item \textbf{Region-level graph representation.} By aggregating micro-contact signals into region-level embeddings, BACE produces compact adjacency matrices that are directly interpretable and suitable for downstream analysis.
    
    \item \textbf{Application to invasive neural recordings.} We demonstrate that BACE yields reproducible and behavior-dependent connectivity estimates in multi-region human LFP data.
\end{itemize}
\section{Related Work}
\textbf{Modeling neural population dynamics.}
Neural population dynamics have been modeled using a wide range of approaches, from classical linear dynamical systems and Gaussian process factor models \citep{dahlhaus2000,yu2008gpfa,gao2016} to modern deep learning methods. These include recurrent neural networks \citep{hochreiter1997long,pandarinath2018}, variational autoencoders \citep{zhou2020pivae}, neural ODEs \citep{kim2021node}, and Transformers adapted for neural time series \citep{NDT,stndt,hosseini2026cross}. Predictive objectives have proven effective for learning latent representations aligned with behavior \citep{sani2021psid,hosseini2025dynamical}. Despite their success in uncovering latent dynamics, these methods are generally not designed to yield the explicit, interpretable, and behavior-conditioned connectivity graphs that are the primary focus of this work.

\textbf{Recurrent models of neural interactions.}
Data-constrained recurrent networks offer another route to
interpreting neural interactions. CURBD fits neuron-level
recurrent networks to multi-region recordings and decomposes
the fitted recurrent inputs into directed currents from source
regions \citep{perich2026curbd}. FacRNN factorizes low-rank
recurrent connectivity into components associated with latent
dynamical groups, encouraging independence between groups
\citep{li2026facrnn}. \citet{beiran2025prediction} study how a
known connectome and partial activity recordings constrain neural
predictions, and show that fitting recurrent weights to recorded
activity alone need not recover the generating connectivity.
BACE provides a complementary approach by directly learning a
compact regional adjacency for each behavioral context while
sharing all other model parameters across contexts.

\textbf{Dynamic graph learning for neural data.}
Graph neural networks (GNNs) have become a common framework for modeling brain activity across modalities. In fMRI, architectures such as STAGIN leverage spatiotemporal attention to learn dynamic representations of functional connectivity for classification tasks \citep{STAGIN}. More recent models combine temporal sequence modeling with graph learning, such as GraphS4mer, which integrates structured state-space models with adaptive graph representations \citep{Graphs4mer}. Similarly, dynamic GNN approaches for EEG, such as EvoBrain, explicitly model time-varying graph structure to capture evolving brain networks \citep{kotoge2025evobrain}. While these methods demonstrate the importance of dynamic graph representations, they are typically evaluated on classification or decoding tasks and often represent connectivity implicitly or in an undirected manner, rather than producing directed adjacency matrices explicitly conditioned on behavioral contexts.

\textbf{Graph structure learning from time series.}
Graph structure learning (GSL) methods aim to jointly infer connectivity and temporal dynamics from multivariate time series \citep{jin2020graph}. In neuroscience, this is particularly relevant because the underlying connectivity is not directly observed. AMAG exemplifies this approach by learning a channel-level adjacency matrix jointly with forecasting neural activity, demonstrating that predictive objectives can recover connectivity from data \citep{AMAG}. More broadly, connections between forecasting and connectivity estimation have been studied in the context of Granger causality, where directed relationships are defined through predictive dependencies \citep{poonia2025exploring}. Recent neural extensions leverage deep sequence models to capture nonlinear dependencies while enforcing sparsity to extract directed connectivity \citep{tank2021neural,poonia2025exploring}. However, these methods typically operate at the level of individual channels and do not explicitly consider behavioral context in modeling.

\textbf{Interpretability of learned neural connectivity.}
Interpretability in graph-based neural models has been studied primarily in the context of classification tasks. Approaches such as BrainGNN identify salient regions using pooling and sparsity regularization \citep{li2021braingnn}, while causal subgraph discovery methods aim to isolate task-relevant connectivity patterns \citep{zheng2024cignn}. Post-hoc explanation methods, including GNNExplainer \citep{ying2019gnnexplainer}, provide edge-level importance scores but do not directly learn interpretable connectivity as a primary model output. In contrast, our approach treats connectivity as an explicit object of interest, producing compact, directed adjacency matrices that can be compared across behavioral contexts.
\section{Methods} 
\subsection{Problem setup}
Many neural recordings measure activity from multiple spatially organized units of analysis, such as anatomical regions, electrode groups, channel groups, or neural populations. We refer to these units as graph nodes. Let $N$ denote the number of nodes, indexed by $i,j\in\{1,\dots,N\}$. Each node $i$ contains $C_i$ recorded channels or contacts, and each sample consists of a recent neural activity window from all nodes.

Each activity window is associated with a behavioral context $k$. The goal is to forecast future neural activity while learning a directed adjacency matrix $\mathbf{A}^{(k)}\in\mathbb{R}^{N\times N}$ for each context. Formally, BACE learns $f_{\theta}(\mathbf{X},k)\mapsto(\hat{\mathbf{Y}},\mathbf{A}^{(k)})$, where $\mathbf{X}$ is the past neural activity window, $\hat{\mathbf{Y}}$ is the predicted future activity, and $\mathbf{A}^{(k)}$ is learned end-to-end as part of the forecasting model. Each adjacency is shared across examples from the same context.

An entry $A_{ij}^{(k)}$ represents the directed influence of source node $j$ on target node $i$ during context $k$. We refer to $\mathbf{A}^{(k)}$ as predictive effective connectivity: a directed connectivity graph learned by forecasting future neural activity within a given behavioral context. These edges should not be interpreted as anatomical connections or as interventional causal effects.

A method for this setting should aggregate fine-grained contacts into interpretable node-level dynamics, learn directed interactions that can vary across behavioral contexts, tie graph estimation to future neural dynamics, and produce compact adjacency matrices that can be inspected and compared across contexts. BACE addresses these goals with the encoder--graph--decoder architecture described below.
\subsection{BACE overview}
\begin{figure}[H]
\vspace{-2mm}
\centering
\includegraphics[width=1.0\textwidth]{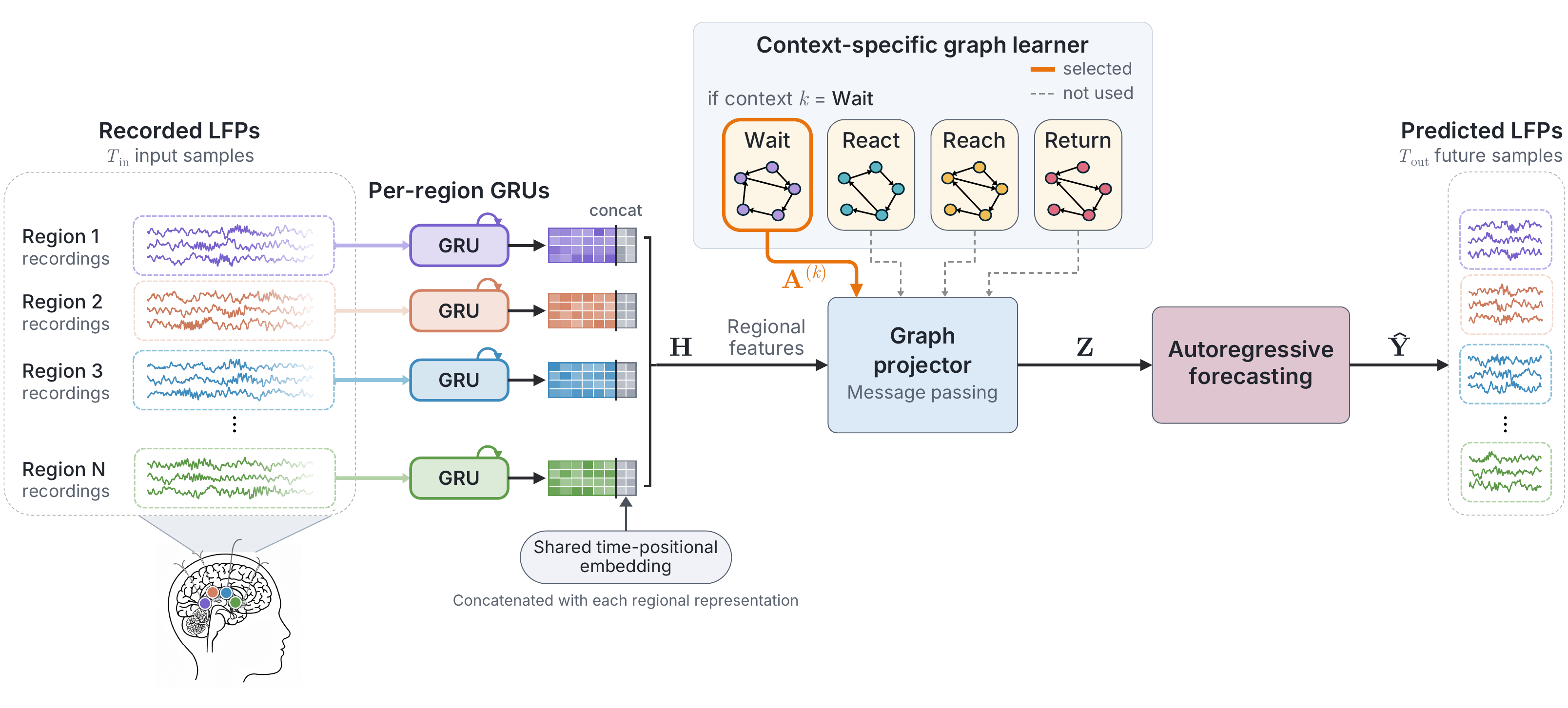}
\caption{\textbf{BACE architecture.} Per-region temporal encoders (independent GRUs) extract local dynamics from multichannel input windows. The behavioral-context label selects a learned directed adjacency $\mathbf{A}^{(k)}$, which mixes regional features in the graph projector and regional predictions during autoregressive feedback. The forecasting head uses temporal attention and a linear readout to produce channel-space residual predictions. This end-to-end design links forecasting directly to explicit, behavior-specific connectivity estimates.}
\label{fig:bace_model}
\vspace{-2mm}
\end{figure}

Figure~\ref{fig:bace_model} summarizes BACE. The model contains four components: regional temporal encoders, a context-specific graph learner, a graph projector, and an autoregressive forecasting head. BACE shares the encoder, graph projector, and forecasting head across behavioral contexts; only the adjacency matrix $\mathbf{A}^{(k)}$ depends on the context label. Thus, context-dependent differences are represented explicitly in the learned connectivity matrices rather than by training separate models for each context.

% \begin{figure}[H]
% \vspace{-2mm}
% \centering
% \includegraphics[width=0.90\textwidth]{figures/Model Figure_new.pdf}
% \caption{\textbf{BACE architecture.} Per-region temporal encoders (independent GRUs) extract local dynamics from multichannel input windows. The behavioral-context label selects a learned directed adjacency $\mathbf{A}^{(k)}$, which mixes regional features in the graph projector and regional predictions during autoregressive feedback. The forecasting head uses temporal attention and a linear readout to produce channel-space residual predictions. This end-to-end design links forecasting directly to explicit, behavior-specific connectivity estimates.}
% \label{fig:bace_model}
% \vspace{-2mm}
% \end{figure}
\paragraph{Regional temporal encoder.}
Neural signals recorded from multiple micro-contacts within the same anatomical region often exhibit shared local dynamics and oscillatory structure \citep{soroushmojdehi2025}. For node $i$, we write the input window as $\mathbf{X}_i\in\mathbb{R}^{C_i\times T_{\mathrm{in}}}$, where $C_i$ is the number of channels and $T_{\mathrm{in}}$ is the input window length in samples. BACE uses a separate GRU encoder \citep{cho2014learning} for each node, $\mathbf{H}_i=\mathrm{GRU}_i(\mathbf{X}_i^\top)$, with time as the recurrent dimension and channels as features at each step, yielding $\mathbf{H}_i\in\mathbb{R}^{T_{\mathrm{in}}\times d_{\mathrm{hidden}}}$. Encoder outputs are layer-normalized and concatenated with a lightweight time-positional embedding, yielding $\mathbf{H}\in\mathbb{R}^{N\times T_{\mathrm{in}}\times d}$. The GRU parameters are not shared across nodes, so within-node temporal dynamics are modeled independently before any cross-node mixing occurs. Inter-node information exchange is mediated only by the learned adjacency $\mathbf{A}^{(k)}$, through graph projection and autoregressive graph feedback. A held-out source-region perturbation analysis supports this architectural property (Appendix~\ref{appendix:ablation_details}).

\paragraph{Context-specific graph learner.}
BACE maintains one directed adjacency matrix $\mathbf{A}^{(k)}$ for each behavioral context $k$. During training and inference, the behavioral context label selects the corresponding adjacency matrix; it does not select a separate encoder, projector, or forecasting head.

Rows of each adjacency matrix are normalized to have unit $\ell_1$ norm, i.e.\ $\sum_j |A_{ij}^{(k)}|=1$ for every $i$, so each row can be interpreted as the relative pattern of incoming predictive influences onto target node $i$. Diagonal entries are fixed to zero. Adjacencies are initialized from Pearson correlations between region-averaged signals computed on training trials only, followed by diagonal masking and row normalization. 

\paragraph{Graph projector.}
Given $\mathbf{A}^{(k)}$, the graph projector introduces directed cross-node coupling. For node $i$ and input time index $\tau$,
{\setlength{\abovedisplayskip}{4pt}
\setlength{\belowdisplayskip}{4pt}
\[
\mathbf{Z}_{i,\tau}
=
W_{\mathrm{self}}\mathbf{H}_{i,\tau}
+
\sum_{j=1}^{N}
A_{ij}^{(k)}W_{\mathrm{neigh}}\mathbf{H}_{j,\tau}.
\]
}
Here, $\mathbf{H}_{i,\tau}\in\mathbb{R}^{d}$ is the encoded representation of node $i$, and $W_{\mathrm{self}}$ and $W_{\mathrm{neigh}}$ are learnable linear maps. The first term preserves self-dynamics, while the second aggregates transformed embeddings from source nodes $j$ according to their directed influence on target node $i$. $W_{\mathrm{neigh}}$ transforms features identically across node pairs and contexts, whereas $A_{ij}^{(k)}$ supplies the explicit source--target-specific and context-specific routing weight. The adjacency mixes temporally encoded representations; it is not a linear state-transition matrix. We denote the activated representation by $\mathbf{Z}^{\mathrm{act}}_{i,\tau}=\mathrm{LeakyReLU}(\mathbf{Z}_{i,\tau})$. These activated regional representations are stacked into decoder memory
$\mathbf{Z}=\mathrm{Stack}_{i,\tau}(\mathbf{Z}^{\mathrm{act}}_{i,\tau})
\in\mathbb{R}^{N\times T_{\mathrm{in}}\times d_{\mathrm{proj}}}$.

\paragraph{Autoregressive forecasting head.}
The forecasting head predicts $T_{\mathrm{out}}$ future samples autoregressively from the decoder memory $\mathbf{Z}$. Let $\mathbf{X}_{\mathrm{last}}\in\mathbb{R}^{N\times C}$ be the last observed sample, where $C$ is the common channel count per node within a participant. We initialize $\hat{\mathbf{Y}}_0=\mathbf{X}_{\mathrm{last}}$ and write $\hat{\mathbf{y}}_{i,t}$ and $\tilde{\mathbf{y}}_{i,t}$ for node $i$'s rows of $\hat{\mathbf{Y}}_t$ and $\tilde{\mathbf{Y}}_t$, respectively. At each forecast step $t=1,\ldots,T_{\mathrm{out}}$, the previous prediction is first refined through graph feedback, $\tilde{\mathbf{Y}}_{t-1}=\hat{\mathbf{Y}}_{t-1}+\alpha_g\mathbf{A}^{(k)}\hat{\mathbf{Y}}_{t-1}$, where $\alpha_g$ sets the feedback strength. For each node, the previous decoder state $\mathbf{s}_{i,t-1}$ then supplies a query over that node's memory only, giving a temporal summary $\mathbf{c}_{i,t}=\sum_{\tau=1}^{T_{\mathrm{in}}}a_{i,t,\tau}W_v\mathbf{Z}^{\mathrm{act}}_{i,\tau}$, with weights $a_{i,t,\tau}$ normalized over input time. Finally, the decoder updates its state, and a linear readout maps it directly to a channel-space residual:
{\setlength{\abovedisplayskip}{4pt}
\setlength{\belowdisplayskip}{4pt}
\[
\mathbf{s}_{i,t}=\mathrm{Dec}\big(\mathbf{s}_{i,t-1},[\mathbf{c}_{i,t};\tilde{\mathbf{y}}_{i,t-1}]\big),
\qquad
\Delta\hat{\mathbf{y}}_{i,t}=W_{\mathrm{out}}\mathbf{s}_{i,t},
\qquad
\hat{\mathbf{y}}_{i,t}=\hat{\mathbf{y}}_{i,t-1}+\Delta\hat{\mathbf{y}}_{i,t}.
\]
}
Each node's output maps only to its own channels, and the loss is computed directly on these channel-space predictions.

\paragraph{Training and optimization.}
BACE is trained end-to-end with a forecasting objective over the next
$T_{\mathrm{out}}$ samples. The objective combines weighted mean-squared prediction error with temporal regularizers:
\[
\begin{aligned}
\mathcal{L}
={}&\frac{1}{T_{\mathrm{out}}}
\sum_{t=1}^{T_{\mathrm{out}}}
w_t\|\hat{\mathbf{Y}}_t-\mathbf{Y}_t\|_2^2
+\lambda_{\mathrm{cont}}
\|\hat{\mathbf{Y}}_1-\mathbf{X}_{\mathrm{last}}\|_2^2\\
&+\lambda_{\mathrm{vel}}
\|\Delta\hat{\mathbf{Y}}_1-\Delta\mathbf{Y}_1\|_2^2
+\lambda_{\mathrm{curv}}
\|\Delta^2\hat{\mathbf{Y}}-\Delta^2\mathbf{Y}\|_2^2.
\end{aligned}
\]
Here, $\mathbf{Y}_t$ is the ground-truth future activity, and $w_t$ weights prediction errors across the forecast horizon. We define $\Delta\hat{\mathbf{Y}}_t=\hat{\mathbf{Y}}_t-\hat{\mathbf{Y}}_{t-1}$, with $\hat{\mathbf{Y}}_0=\mathbf{Y}_0=\mathbf{X}_{\mathrm{last}}$, and define target differences analogously. The continuity term penalizes the first predicted jump from the observed window, the velocity term matches the first predicted increment to the observed increment, and the curvature term matches second differences over the forecast horizon.

All model parameters, including the context-specific adjacency matrices, are optimized jointly with Adam. We use gradient clipping and early stopping based on validation loss, retaining the checkpoint with the lowest validation loss for test evaluation. Additional model and optimization details are provided in
Appendix~\ref{appendix:experimental_details}.

\section{Experiments and results}
\subsection{Synthetic validation of graph recovery under known ground truth}
\begin{figure}[htbp]\center{\includegraphics[width=1\textwidth,clip=false] {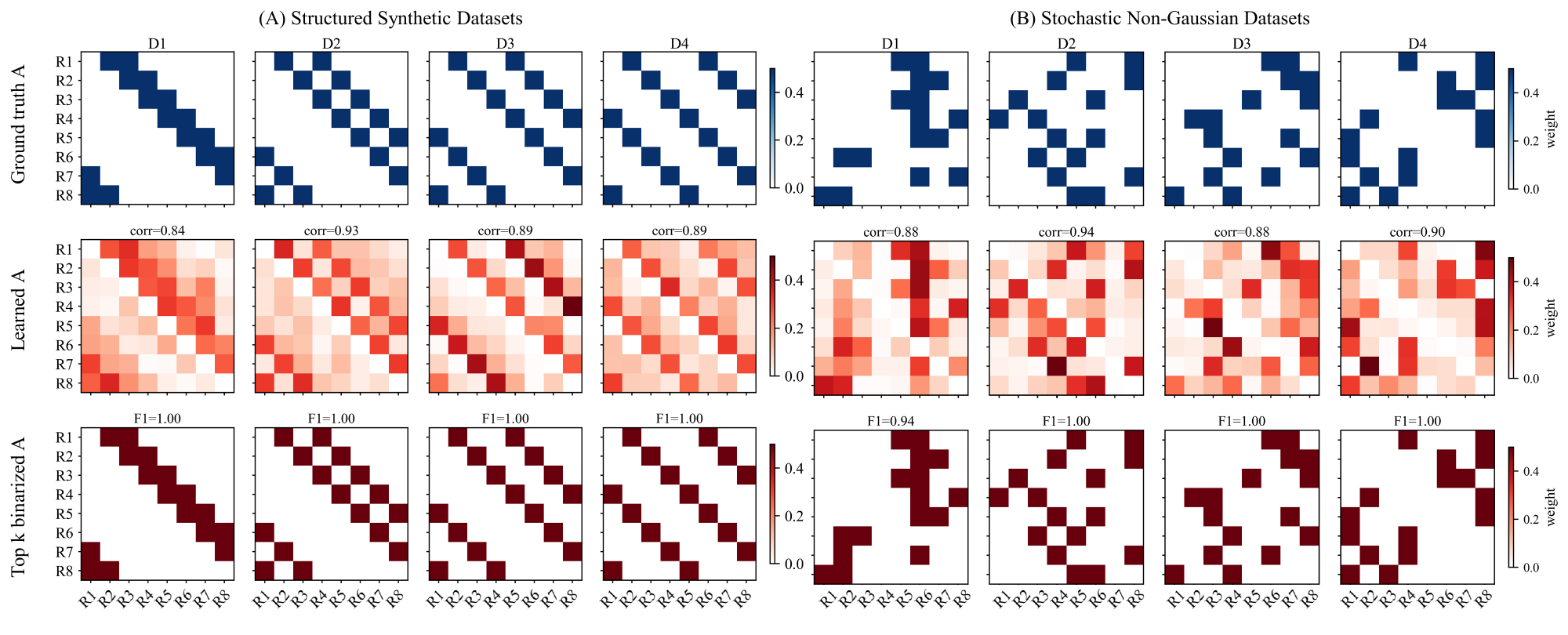}}
\caption{\textbf{Adjacency recovery on synthetic datasets.}
(\textbf{A}) Structured suite; (\textbf{B}) stochastic non-Gaussian suite.
Columns correspond to the four dataset regimes $\mathcal{D}_1$--$\mathcal{D}_4$.
Rows show, from top to bottom: the ground-truth adjacency $\mathbf{A}$, BACE's learned adjacency $\hat{\mathbf{A}}$, and the top-$k$ binarization of $\hat{\mathbf{A}}$, with $k$ set to the ground-truth row degree ($k{=}2$).
Values above the learned-adjacency panels report Pearson correlation with the ground-truth weights; values above the binarized panels report \textsc{F1@}$k_{\text{row}}$.}
\label{fig:synth_recovery}
\end{figure}

We use controlled synthetic time series to test whether BACE can recover a known directed interaction graph from a forecasting objective alone. These datasets are designed to match the scale of our neural recordings (small region-level graphs, hundreds of trials, and short segmented windows) but not the exact anatomical configuration or biological complexity of intracranial LFP. This experiment isolates the graph-learning mechanism: can the adjacency learned by BACE recover the known directed edge structure used to generate the data, even though the model is trained only to forecast future signals?
\paragraph{Dataset design.} We generated two synthetic suites, each containing four regimes $\mathcal{D}_1$--$\mathcal{D}_4$ with distinct ground-truth adjacency matrices. In the structured suite, each regime used a sparse directed graph with the same row degree but different edge placements, producing visually distinct connectivity patterns. In the stochastic non-Gaussian suite, time series were generated from linear non-Gaussian dynamics following prior work on structure learning from neural time series \citep{AMAG,biswas2022tapc,biswas2022comparative,song2009time}.
The governing dynamics take the form
$\mathbf{X}_t = \mathbf{X}_{t-1}
+ G(\mathbf{X}_{t-1}, \mathbf{A}) + \boldsymbol{\mu}_t$,
where
$G(\mathbf{X}_{t-1}, \mathbf{A})
= (-\lambda\mathbf{I}+\gamma\mathbf{A})\mathbf{X}_{t-1}$
combines linear leak and adjacency-mediated coupling.
The noise $\boldsymbol{\mu}_t$ is autoregressive with uniform
innovations, making it colored and non-Gaussian.
Since this suite uses linear dynamics, we evaluate VAR/Granger
as a matched baseline.
\paragraph{Synthetic validation.}
BACE was trained using only the forecasting loss; the ground-truth adjacencies were used only for evaluation. Graph recovery was assessed with two metrics: (i) \textsc{F1@}$k_{\text{row}}$, which measures whether the top-$k$ entries in each row match the ground-truth nonzero edges, and (ii) Pearson correlation between learned and ground-truth edge weights. Across the structured suite, BACE achieved perfect support recovery (mean \textsc{F1@}$k_{\text{row}}{=}1.00$) with mean correlation $0.89$. Across the stochastic non-Gaussian suite, recovery remained strong, with mean \textsc{F1@}$k_{\text{row}}{=}0.98$ and mean correlation $0.90$ (per-regime values in Fig.~\ref{fig:synth_recovery}). As a positive-control sanity check, a matched VAR/Granger estimator nearly perfectly recovered the stochastic-suite graphs (\textsc{F1@}$k_{\text{row}}{=}1.00$, correlation $0.99$). These results show that BACE can recover planted directed
structure in the evaluated synthetic systems.

\paragraph{Graph recovery under nonlinear dynamics.}
To test whether recovery extends to nonlinear interactions,
we additionally evaluated an eight-node system with four
context-specific graphs, two planted parents per target, and
centered $\tanh^2$ inter-regional coupling
(Appendix~\ref{appendix:nonlinear_synthetic}).
Across three runs, BACE achieved
\textsc{F1@}$k_{\text{row}}=1.000\pm0.000$ ($k=2$),
compared with $0.474\pm0.018$ for context-specific VAR(10),
whose lag range included the true structural lag.
Pearson correlation between absolute inferred and ground-truth
edge weights was $0.989\pm0.002$ for BACE versus
$0.381\pm0.095$ for VAR(10).
These results extend graph recovery to a nonlinear setting
in which BACE outperforms the tested linear comparator.
\subsection{Evaluation on human LFP data}
\paragraph{Dataset and preprocessing.}
We evaluated BACE on 3 human intracranial LFP datasets collected during DBS treatment as part of clinical care. Recordings were acquired during a cued reaching task and segmented into 4 behavioral contexts: Wait, the 400~ms pre-cue baseline; React, the 400~ms after the go cue before movement; Reach, the forward reach to the target; and Return, the movement back to the home button. Contacts were grouped into anatomically defined nodes, yielding 8--10 nodes per participant with 8--10 retained channels per node after cleaning. BACE was trained separately for each participant to accommodate differences in electrode coverage \citep[see][for cross-subject aggregation]{javadzadeh2025functionalembed}. Node indices are therefore not homologous across participants.

Signals were filtered from 1--50~Hz, common-average referenced, downsampled to 1~kHz, and cleaned by removing noisy channels and outlier trials. Each 400~ms context window was divided into sliding forecasting examples with $T_{\mathrm{in}}=100$, $T_{\mathrm{out}}=20$, and stride 20. Within each behavioral context, trials were divided into 70\% training, 15\% validation, and 15\% test sets before sliding windows were generated, preventing leakage through overlapping windows; the same split was used for BACE, ablations, and baselines. For each participant, all analyzed trials were recorded within a single session, so reported results reflect generalization to held-out trials within that session. Participant-level recording and preprocessing details are provided in Appendix~\ref{appendix:data_details}.
\noindent
\paragraph{Benchmark methods.}
To contextualize BACE's forecasting performance, we compare against a targeted set of baselines spanning the main modeling families relevant to multivariate neural time-series prediction: classical linear dynamics, recurrent neural networks, Transformer-based sequence models, and graph-based neural forecasting. As a classical effective-connectivity reference, we include a VAR/Granger model. From the recurrent family, we include a standard LSTM~\citep{hochreiter1997long}, together with a context-aware LSTM variant that receives the same behavioral context labels as BACE. From the Transformer family, we evaluate the Neural Data Transformer (NDT)~\citep{NDT}, designed for neural sequence modeling with masked reconstruction objectives. For graph-based methods, we include DCRNN~\citep{DCRNN}, which integrates graph diffusion over a fixed adjacency into recurrent updates; GWNet~\citep{GWNET}, which combines dilated causal convolutions with adaptive graph learning; and AMAG~\citep{AMAG}, which adaptively learns channel-level graphs for neural forecasting. We also include a context-aware AMAG variant that receives the same behavioral context labels as BACE. These baselines provide reference points for predictive accuracy, while BACE is designed to additionally return compact, behavior-indexed directed region-level graphs for downstream connectivity analysis.
\paragraph{Experimental setup.}
All neural models were trained separately for each participant and evaluated using matched participant-specific train/validation/test splits. The VAR/Granger baseline was evaluated across three different trial splits. Values report mean $\pm$ standard deviation across runs. Training used early stopping based on validation loss, and the checkpoint with the best validation performance was used for test evaluation. Hyperparameters and implementation details for BACE and all baselines
are provided in Appendices~\ref{appendix:experimental_details}
and~\ref{appendix:baseline_details}. We report mean squared error (MSE), coefficient of determination ($R^2$), and Pearson correlation (Corr): MSE measures absolute prediction error, $R^2$ quantifies explained variance, and Corr measures temporal alignment between predicted and ground-truth activity.

\begin{figure}[H]
\centering
\includegraphics[width=0.9\textwidth]{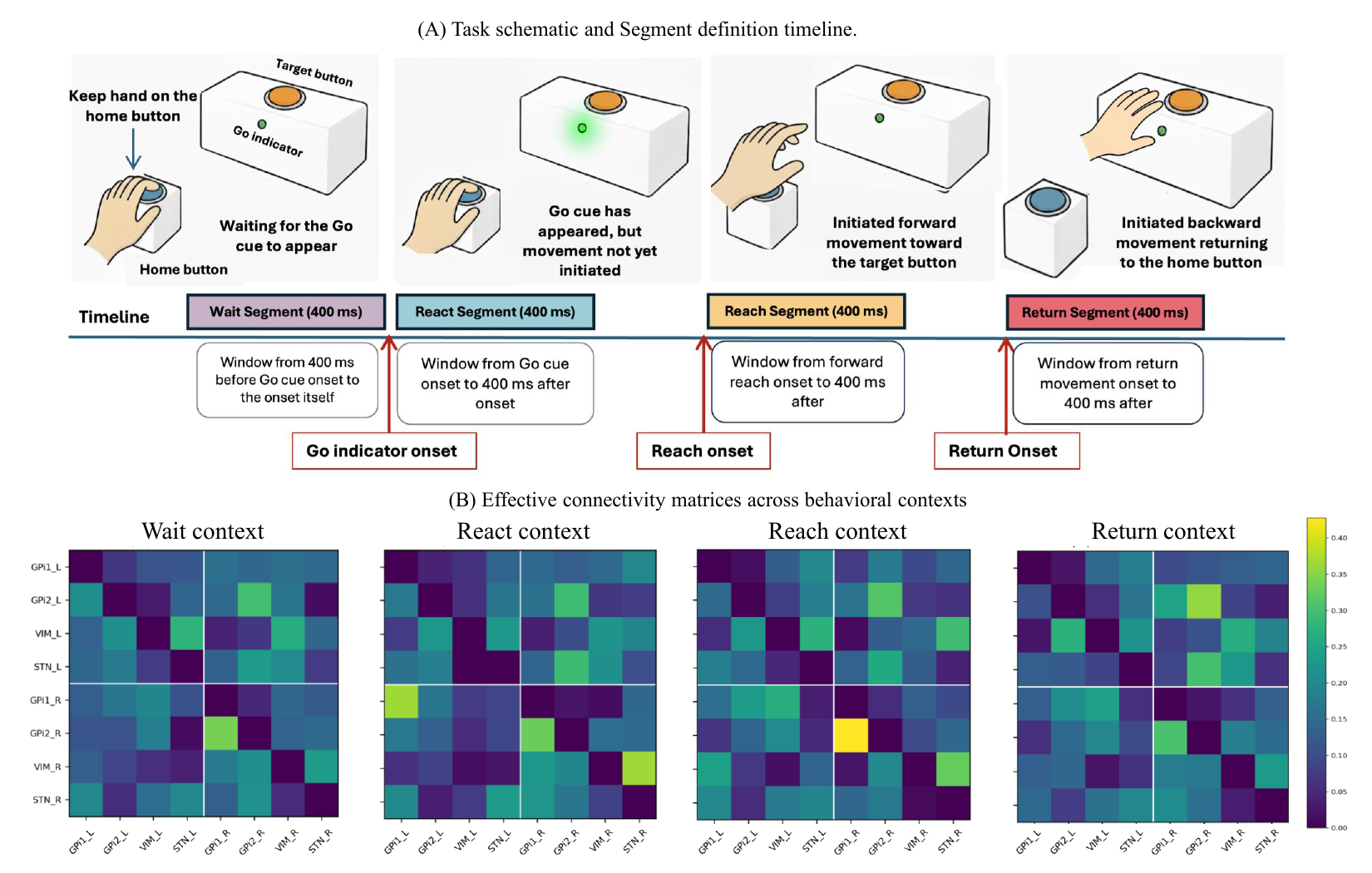}
\caption{(\textbf{A}) Task schematic and behavioral segmentation timeline. Each trial comprised 4 stages with a 400~ms analysis window: Wait ($-400$~ms to go cue onset), React (0--400~ms after go cue onset), Reach (0--400~ms after forward reach onset), and Return (0--400~ms after return onset). 
(\textbf{B}) Representative context-specific $N\times N$ predictive effective connectivity matrices estimated by BACE for one participant. Matrix $\mathbf{A}^{(k)}_{ij}$ denotes directed influence from source node $j$ to target node $i$. Node labels GPi1 and GPi2 denote anterior and
posterior GPi, respectively; \_L/\_R indicate hemisphere.}
\label{fig:lfp_connectivity}
\vspace{-3mm}
\end{figure}
\paragraph{Model performance.} 
Table~\ref{tab:benchmarks} summarizes 20-step forecasting performance across the 3 human intracranial LFP datasets. The neural models were evaluated on the same participant-specific train/validation/test splits, and values report mean $\pm$ standard deviation across runs. BACE achieved the lowest MSE and highest $R^2$ and Corr for all 3 participants, showing that its forecasting performance was consistent across different datasets.

Among non-graph models, LSTM and NDT provided reasonable forecasts but were consistently outperformed by graph-based methods. VAR/Granger was a strong classical baseline, showing that LFP forecasting contains substantial linear predictive structure. However, learned graph-based models improved further, suggesting that modeling interactions across recorded channels or regions is beneficial. Among the graph-based baselines, GWNet and AMAG were the strongest competitors, with AMAG providing the closest comparison because it also learns adaptive graphs for neural forecasting. Because BACE uses the behavioral context only to select the graph used in the graph-mixing layer, we added context-aware baselines that received the behavioral context labels for a more balanced comparison. The context-aware variants improved over their original versions, especially for AMAG, indicating that behavioral context helps prediction. Nevertheless, BACE achieved the best performance across all 3 datasets. Overall, BACE provides strong forecasting accuracy while also returning compact context-specific graphs that can be used to study directed connectivity between recorded regions.

\begin{table}[H]
\centering
\vspace{-3mm}
\caption{20-step forecasting results across three human intracranial LFP datasets. Values report mean $\pm$ standard deviation across three runs. Ctx denotes context-aware variants. Best results are shown in bold.}
\label{tab:benchmarks}

\vspace{1mm}
\scriptsize
\setlength{\tabcolsep}{2.2pt}
\renewcommand{\arraystretch}{0.92}
\resizebox{\textwidth}{!}{
\begin{tabular}{lccccccccc}
\toprule
& \multicolumn{3}{c}{P1} & \multicolumn{3}{c}{P2} & \multicolumn{3}{c}{P3} \\
\cmidrule(lr){2-4} \cmidrule(lr){5-7} \cmidrule(lr){8-10}
Model
& $R^2 \uparrow$ & Corr $\uparrow$ & MSE $\downarrow$
& $R^2 \uparrow$ & Corr $\uparrow$ & MSE $\downarrow$
& $R^2 \uparrow$ & Corr $\uparrow$ & MSE $\downarrow$ \\
\midrule

LSTM
& 0.779 $\pm$ 4e$^{-4}$ & 0.886 $\pm$ 5e$^{-4}$ & 0.1645 $\pm$ 3e$^{-4}$
& 0.748 $\pm$ 1e$^{-3}$ & 0.871 $\pm$ 1e$^{-3}$ & 0.1739 $\pm$ 1e$^{-3}$
& 0.636 $\pm$ 3e$^{-3}$ & 0.799 $\pm$ 3e$^{-3}$ & 0.3014 $\pm$ 3e$^{-3}$ \\

Ctx-LSTM
& 0.781 $\pm$ 3e$^{-3}$ & 0.887 $\pm$ 1e$^{-3}$ & 0.1631 $\pm$ 2e$^{-3}$
& 0.751 $\pm$ 3e$^{-3}$ & 0.873 $\pm$ 3e$^{-3}$ & 0.1717 $\pm$ 2e$^{-3}$
& 0.638 $\pm$ 2e$^{-3}$ & 0.801 $\pm$ 1e$^{-3}$ & 0.2998 $\pm$ 2e$^{-3}$ \\

NDT
& 0.786 $\pm$ 1e$^{-3}$ & 0.889 $\pm$ 6e$^{-4}$ & 0.1594 $\pm$ 1e$^{-3}$
& 0.760 $\pm$ 2e$^{-3}$ & 0.877 $\pm$ 5e$^{-4}$ & 0.1656 $\pm$ 2e$^{-3}$
& 0.649 $\pm$ 3e$^{-4}$ & 0.807 $\pm$ 1e$^{-4}$ & 0.2908 $\pm$ 2e$^{-4}$ \\

VAR/Granger
& 0.837 $\pm$ 4e$^{-3}$ & 0.916 $\pm$ 2e$^{-3}$ & 0.1209 $\pm$ 3e$^{-3}$
& 0.825 $\pm$ 4e$^{-3}$ & 0.909 $\pm$ 2e$^{-3}$ & 0.1207 $\pm$ 3e$^{-3}$
& 0.773 $\pm$ 1e$^{-2}$ & 0.879 $\pm$ 8e$^{-3}$ & 0.1812 $\pm$ 1e$^{-2}$ \\

DCRNN
& 0.842 $\pm$ 4e$^{-2}$ & 0.918 $\pm$ 2e$^{-2}$ & 0.1179 $\pm$ 3e$^{-2}$
& 0.837 $\pm$ 8e$^{-3}$ & 0.918 $\pm$ 6e$^{-3}$ & 0.1123 $\pm$ 6e$^{-3}$
& 0.685 $\pm$ 2e$^{-2}$ & 0.835 $\pm$ 2e$^{-2}$ & 0.2612 $\pm$ 2e$^{-2}$ \\

GWNet
& 0.859 $\pm$ 8e$^{-4}$ & 0.940 $\pm$ 4e$^{-3}$ & 0.1051 $\pm$ 6e$^{-4}$
& 0.829 $\pm$ 1e$^{-3}$ & 0.930 $\pm$ 4e$^{-3}$ & 0.1184 $\pm$ 7e$^{-4}$
& 0.793 $\pm$ 1e$^{-3}$ & 0.914 $\pm$ 4e$^{-3}$ & 0.1715 $\pm$ 1e$^{-3}$ \\

AMAG
& 0.861 $\pm$ 7e$^{-3}$ & 0.928 $\pm$ 4e$^{-3}$ & 0.1030 $\pm$ 5e$^{-3}$
& 0.832 $\pm$ 8e$^{-3}$ & 0.929 $\pm$ 5e$^{-3}$ & 0.1160 $\pm$ 6e$^{-3}$
& 0.797 $\pm$ 4e$^{-3}$ & 0.899 $\pm$ 4e$^{-3}$ & 0.1680 $\pm$ 4e$^{-3}$ \\

Ctx-AMAG
& 0.880 $\pm$ 5e$^{-3}$ & 0.939 $\pm$ 3e$^{-3}$ & 0.0890 $\pm$ 4e$^{-3}$
& 0.841 $\pm$ 1e$^{-2}$ & 0.929 $\pm$ 6e$^{-3}$ & 0.1100 $\pm$ 8e$^{-3}$
& 0.814 $\pm$ 2e$^{-2}$ & 0.904 $\pm$ 1e$^{-2}$ & 0.1539 $\pm$ 2e$^{-2}$ \\

BACE
& \textbf{0.894 $\pm$ 3e$^{-4}$} & \textbf{0.946 $\pm$ 1e$^{-4}$} & \textbf{0.0787 $\pm$ 2e$^{-4}$}
& \textbf{0.874 $\pm$ 3e$^{-4}$} & \textbf{0.935 $\pm$ 1e$^{-4}$} & \textbf{0.0867 $\pm$ 2e$^{-4}$}
& \textbf{0.845 $\pm$ 6e$^{-4}$} & \textbf{0.919 $\pm$ 3e$^{-4}$} & \textbf{0.1285 $\pm$ 5e$^{-4}$} \\

\bottomrule
\end{tabular}
}
\vspace{-2mm}
\end{table}

\paragraph{Ablation: effect of behavior-adaptive graph structure.}
To assess the contributions of graph structure and behavioral-context
conditioning, we evaluated four BACE variants across all three
participants: NoGraph removes graph mixing and graph feedback;
Ctx-NoGraph adds the behavioral context label to the NoGraph
forecasting decoder; SingleGraph learns one adjacency matrix shared
across behavioral contexts; and Full BACE learns one
$\mathbf{A}^{(k)}$ per context while sharing the encoder, graph
projector, and forecasting head across contexts. All variants retain
the same regional temporal backbone and forecasting objective and
use matched training settings and identical data splits.

As shown in Table~\ref{tab:ablation_graph}, Full BACE achieved the lowest mean MSE for all three participants, improving over both NoGraph and SingleGraph. Full BACE also consistently outperformed the context-aware graph-free control, Ctx-NoGraph, supporting the contribution of graph-mediated inter-regional information beyond behavioral-context conditioning alone. Additional ablation details and a complementary source-region perturbation analysis are provided in Appendix~\ref{appendix:ablation_details}.

\begin{table}[H]
\centering
\caption{Ablation of graph structure across 3 participants. Ctx-NoGraph receives the behavioral context label but has no graph mixing or graph feedback. Values report mean $\pm$ standard deviation over three seeds.}
\label{tab:ablation_graph}
\vspace{1mm}
\setlength{\tabcolsep}{3pt}
\renewcommand{\arraystretch}{0.92}
\scriptsize
\resizebox{\textwidth}{!}{%
\begin{tabular}{lccccccccc}
\toprule
& \multicolumn{3}{c}{P1}
& \multicolumn{3}{c}{P2}
& \multicolumn{3}{c}{P3} \\
\cmidrule(lr){2-4}
\cmidrule(lr){5-7}
\cmidrule(lr){8-10}
Model
& $R^2 \uparrow$ & Corr $\uparrow$ & MSE $\downarrow$
& $R^2 \uparrow$ & Corr $\uparrow$ & MSE $\downarrow$
& $R^2 \uparrow$ & Corr $\uparrow$ & MSE $\downarrow$ \\
\midrule

NoGraph
& 0.862 $\pm$ 3e$^{-4}$ & 0.928 $\pm$ 2e$^{-4}$ & 0.103 $\pm$ 2e$^{-4}$
& 0.855 $\pm$ 3e$^{-4}$ & 0.924 $\pm$ 1e$^{-4}$ & 0.100 $\pm$ 2e$^{-4}$
& 0.765 $\pm$ 6e$^{-4}$ & 0.874 $\pm$ 3e$^{-4}$ & 0.195 $\pm$ 5e$^{-4}$ \\

Ctx-NoGraph
& 0.882 $\pm$ 4e$^{-4}$ & 0.939 $\pm$ 2e$^{-4}$ & 0.088 $\pm$ 3e$^{-4}$
& 0.865 $\pm$ 8e$^{-4}$ & 0.931 $\pm$ 4e$^{-4}$ & 0.093 $\pm$ 6e$^{-4}$
& 0.827 $\pm$ 2e$^{-3}$ & 0.910 $\pm$ 1e$^{-3}$ & 0.143 $\pm$ 2e$^{-3}$ \\

SingleGraph
& 0.886 $\pm$ 3e$^{-4}$ & 0.941 $\pm$ 2e$^{-4}$ & 0.085 $\pm$ 3e$^{-4}$
& 0.862 $\pm$ 4e$^{-4}$ & 0.928 $\pm$ 1e$^{-4}$ & 0.095 $\pm$ 3e$^{-4}$
& 0.831 $\pm$ 5e$^{-4}$ & 0.911 $\pm$ 3e$^{-4}$ & 0.140 $\pm$ 4e$^{-4}$ \\

Full BACE
& \textbf{0.894 $\pm$ 3e$^{-4}$} & \textbf{0.946 $\pm$ 1e$^{-4}$} & \textbf{0.079 $\pm$ 2e$^{-4}$}
& \textbf{0.874 $\pm$ 3e$^{-4}$} & \textbf{0.935 $\pm$ 1e$^{-4}$} & \textbf{0.087 $\pm$ 2e$^{-4}$}
& \textbf{0.845 $\pm$ 6e$^{-4}$} & \textbf{0.919 $\pm$ 3e$^{-4}$} & \textbf{0.129 $\pm$ 5e$^{-4}$} \\

\bottomrule
\end{tabular}%
}
\vspace{-2mm}
\end{table}

\paragraph{Computational complexity.}
BACE is lightweight relative to the neural forecasting baselines, with 0.13M trainable parameters and 0.20 GB peak GPU memory in our profiling setup, while also achieving the best forecasting results in Table~\ref{tab:benchmarks}. Full parameter counts, memory measurements, runtime details, and hardware specifications are provided in Appendix~\ref{appendix:compute}.
\subsection{Analysis of learned connectivity}
\paragraph{Reproducibility.} 
We first asked whether BACE's learned connectivity graphs were stable across resampled subsets of trials. For each participant and behavioral context, we repeated split-half resampling and compared the resulting absolute adjacency matrices $|\mathbf{A}^{(k)}|$. Across the 3 participants and 4 behavioral contexts, split-half correlations ranged from $r=0.83$ to $0.91$, corresponding to Spearman--Brown corrected repeatability values of $0.91$ to $0.95$. These results show that BACE learns context-specific graphs that are reproducible across independent subsets of trials. Full participant-level results are provided in Appendix~\ref{appendix:stats}.

\begin{figure}[!t]
\centering
\includegraphics[width=\textwidth]{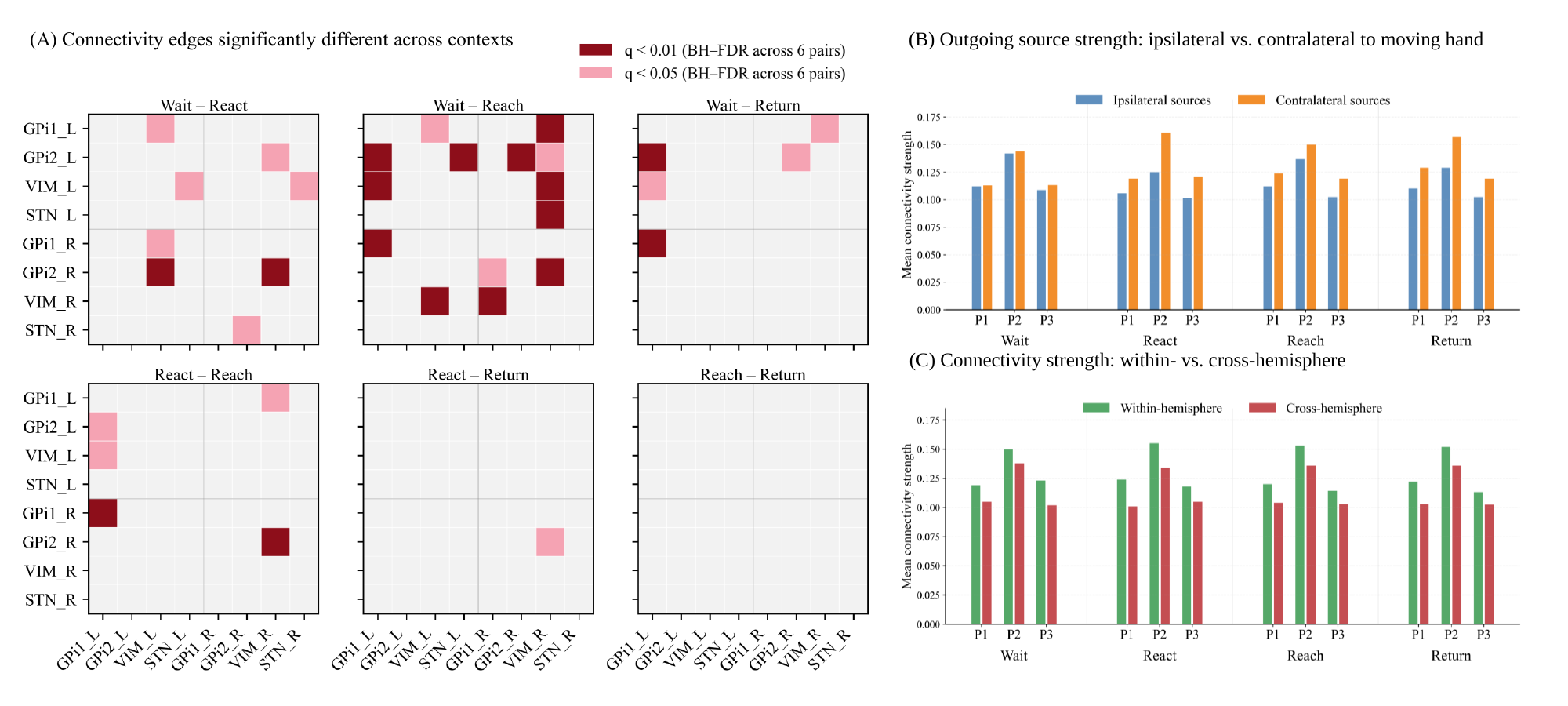}
\caption{
\textbf{Behavior-specific connectivity analyses.}
(\textbf{A}) Representative participant showing edges with significant differences across behavioral contexts, based on bootstrap confidence intervals and Benjamini--Hochberg FDR correction.
(\textbf{B}) Across-participant average outgoing connectivity strength from ipsilateral vs.~contralateral hemisphere sources, defined relative to the moving hand.
(\textbf{C}) Across-participant average within-hemisphere vs.~cross-hemisphere connectivity strength.
}
\label{fig:group_results}
\end{figure}

\paragraph{Edge reliability and context-dependent changes.}  
We next tested whether the learned directed edges differed across behavioral contexts. For each participant and each adjacency matrix $\mathbf{A}^{(k)}$, we used nonparametric bootstrap resampling over trials to estimate confidence intervals for edge weights. We then tested pairwise differences between behavioral contexts with FDR correction \citep{benjamini1995controlling}. Across all 3 participants, several directed edges showed reliable context-dependent changes. Fig.~\ref{fig:group_results}A shows one representative participant. These results indicate that BACE does not return a single fixed connectivity pattern, but instead learns graphs that vary with task segment.
\paragraph{Spatial organization.}  
We next examined whether the learned graphs showed interpretable spatial structure across participants. We first compared within-hemisphere and cross-hemisphere edges. Within-hemisphere edges connect nodes in the same hemisphere, whereas cross-hemisphere edges connect nodes across hemispheres. BACE estimated stronger within-hemisphere than cross-hemisphere connectivity across behavioral contexts and participants (Fig.~\ref{fig:group_results}C), consistent with the general observation that brain networks often show stronger local or intra-hemispheric coupling than inter-hemispheric coupling \citep{bullmore2012,sporns2016}. We also compared outgoing source strength from hemispheres ipsilateral and contralateral to the moving hand. Across participants, contralateral sources showed stronger outgoing influence than ipsilateral sources during React, Reach, and Return, while Wait was more balanced (Fig.~\ref{fig:group_results}B). This pattern is consistent with contralateral motor organization \citep{shibasaki2006bereitschaftspotential,vingerhoets2014}. Together, these analyses suggest that BACE produces directed predictive graphs that are not only accurate for forecasting, but also structured in ways that align with known spatial organization of motor networks.

\section{Discussion}
This work introduced \textsc{BACE}, a framework for estimating behavior-adapted connectivity from multi-region neural recordings. \textsc{BACE} couples forecasting with graph estimation, learning compact, directed inter-regional graphs whose parameters encode predictive interactions. By operating at the region level, the model consolidates many micro-contacts into interpretable embeddings and then learns connectivity among regions. The resulting adjacency matrices are behavior-specific and interpretable, supporting network analyses that can be inspected by domain experts.

In synthetic experiments, BACE recovered ground-truth connectivity while trained with a forecasting objective, suggesting that prediction-based objectives can recover connectivity structure from multivariate time series. Applied to 3 human subcortical LFP datasets during a cued reaching task, the model produced one connectivity matrix per behavioral context for each participant. Comparisons revealed behavior-dependent modulation of inter-regional influence, suggesting that deep-brain networks change their directed connectivity across task segments. The learned graphs also reflected interpretable spatial structure: within-hemisphere edges were stronger than cross-hemisphere edges, consistent with wiring-cost and modularity constraints \citep{bullmore2012,sporns2016}.
Methodologically, several design choices contribute to interpretability: independent regional encoders prevent cross-region leakage prior to graph mixing, context-specific adjacency matrices isolate behavior-dependent graph structure, and lightweight temporal regularizers stabilize multi-step predictions.

The current implementation of BACE uses experimentally defined behavioral contexts to select the corresponding adjacency matrix, while keeping the encoder and decoder shared across contexts; future extensions could relax this requirement by learning contexts in a weakly supervised or continuous manner. The estimated graphs capture predictive effective connectivity rather than anatomical or interventional causal effects. Future work should evaluate BACE across additional recording sessions, larger cohorts, additional tasks, physiology-informed screening \citep{farahani2026beyondahi}, and other recording modalities, and could incorporate anatomical priors or multimodal behavioral signals such as EMG and video. Clinically, behavior-adapted connectivity estimates may provide a foundation for hypothesis generation and exploratory studies of brain-network analysis, but any translational use will require substantial validation. \section*{Reproducibility statement}
Code for model implementation, training, and evaluation is available at 
\url{https://github.com/mehrnazasadi95/BACE}.
\begin{ack}
This work was supported by the Once Upon a Time Foundation
through the Raynor Cerebellum Project.
\end{ack}
% \url{https://github.com/anonymresearch483/neurips-anonymous-code}.
\bibliographystyle{plainnat}
\bibliography{references}

%%%%%%%%%%%%%%%%%%%%%%%%%%%%%%%%%%%%%%%%%%%%%%%%%%%%%%%%%%%%

\appendix

\section{Method Details}
\subsection{Model and Training Details}
\label{appendix:experimental_details}

\paragraph{Adjacency parameterization and initialization.}
For each behavioral context $k$, BACE learns a directed adjacency matrix
$\mathbf{A}^{(k)}\in\mathbb{R}^{N\times N}$. During training, the diagonal entries are fixed to zero,
and each row is normalized to unit $\ell_1$ norm,
\[
\sum_j |A^{(k)}_{ij}| = 1,
\]
so that each row represents the relative pattern of incoming predictive influence onto target node
$i$. The same masking and normalization are used for all behavioral contexts. Adjacency matrices
are initialized from region-wise Pearson correlations computed using training trials only. Correlations
are computed between region-averaged signals, the diagonal is set to zero, and the resulting matrices
are row-normalized before training. Validation and test trials are not used for graph initialization.

\paragraph{Time-positional embedding.}
After each region-specific GRU encoder, hidden states are layer-normalized and concatenated with
a lightweight time-positional embedding. Normalized time indices
$\tau/T_{\mathrm{in}}\in[0,1]$ are passed through a linear projection followed by a $\tanh$
activation. This gives the forecasting head access to the relative position of each sample within the
input window while preserving the separation between regions before graph mixing.

\paragraph{Temporal attention in the forecasting head.}
Attention is computed over input time separately within each node.
The previous decoder state supplies the query
$\mathbf{q}_{i,t}=W_q\mathbf{s}_{i,t-1}$, while the activated
graph-projected representations supply keys
$\mathbf{k}_{i,\tau}=W_k\mathbf{Z}^{\mathrm{act}}_{i,\tau}$
and values
$\mathbf{v}_{i,\tau}=W_v\mathbf{Z}^{\mathrm{act}}_{i,\tau}$:
\[
a_{i,t,\tau}
=
\frac{\exp(\beta\,\mathbf{q}_{i,t}^{\top}\mathbf{k}_{i,\tau})}
{\sum_{\tau'=1}^{T_{\mathrm{in}}}
\exp(\beta\,\mathbf{q}_{i,t}^{\top}\mathbf{k}_{i,\tau'})},
\qquad
\mathbf{c}_{i,t}
=
\sum_{\tau=1}^{T_{\mathrm{in}}}
a_{i,t,\tau}\mathbf{v}_{i,\tau}.
\]
Here, $\beta$ is a learnable temperature. The temporal summary
is concatenated with the adjacency-refined previous prediction
to update the regional decoder state and produce the next
channel-space residual forecast.

\paragraph{Temporal differences in the loss.}
The continuity, velocity, and curvature terms in the training objective are computed along the
forecast horizon. We define
\[
\Delta\hat{\mathbf{Y}}_t = \hat{\mathbf{Y}}_t-\hat{\mathbf{Y}}_{t-1},
\qquad
\Delta\mathbf{Y}_t = \mathbf{Y}_t-\mathbf{Y}_{t-1},
\]
with $\hat{\mathbf{Y}}_0=\mathbf{X}_{\mathrm{last}}$ and $\mathbf{Y}_0$ defined analogously. Second-order
differences are computed as
\[
\Delta^2\hat{\mathbf{Y}}_t
=
\Delta\hat{\mathbf{Y}}_t-\Delta\hat{\mathbf{Y}}_{t-1},
\qquad
\Delta^2\mathbf{Y}_t
=
\Delta\mathbf{Y}_t-\Delta\mathbf{Y}_{t-1}.
\]
The horizon weights $w_t$ are log-spaced, with larger weights assigned to later forecast steps.

\paragraph{Optimization.}
All BACE experiments used Adam with learning rate $10^{-3}$, weight decay $10^{-4}$, and gradient
clipping at $1.0$. Training used early stopping based on validation loss, and the checkpoint with
the lowest validation loss was used for test evaluation. Normalization statistics and graph initialization were computed
from training trials only. Held-out test trials were
used only for final evaluation.

\paragraph{Interpretation of the learned adjacency.}
\label{appendix:adj_interpretation}
Because $\mathbf{A}^{(k)}$ is optimized jointly with the regional encoders, the graph projector, and the forecasting head, it is not uniquely identifiable: different parameterizations of the surrounding model could, in principle, yield similar forecasts with different graphs. We therefore interpret $\mathbf{A}^{(k)}$ as a reproducible, context-dependent directed predictive connectivity map within BACE's parameterization, rather than as a uniquely determined causal or anatomical graph. Within this parameterization, the shared projection $W_{\mathrm{neigh}}$ transforms source-node features identically for all node pairs and contexts, so source--target-specific and context-specific routing is carried only by $\mathbf{A}^{(k)}$. Edge magnitudes reflect relative incoming predictive influence after row-wise $\ell_1$ normalization and should not be interpreted as absolute coupling strength or as excitatory or inhibitory synaptic sign.
\subsection{Nonlinear synthetic graph-recovery experiment}
\label{appendix:nonlinear_synthetic}

\paragraph{Data generation.}
We simulated an eight-node system with four context-specific directed graphs $\mathbf{A}^{(k)}$, in which each target node had two planted incoming parents and no self-edges. Each node was represented by a single signal, with dynamics
\[
x_i(t+1) = 0.35\,x_i(t) + 0.50\sum_{j} A^{(k)}_{ij}\left[\tanh^2\!\big(3x_j(t)\big) - \mu_{\phi}\right] + \eta_i(t),
\]
where $\mu_{\phi}$ is a centering constant and $\eta_i(t)$ is an innovation term. Each context contained 326 trials. Ground-truth graphs were used only for evaluation.

\paragraph{Training and comparison.}
BACE was trained using only the forecasting objective, with $T_{\mathrm{in}}=100$ and $T_{\mathrm{out}}=20$. The comparator was a context-specific VAR(10), whose lag range included the true structural lag; its directed edge scores aggregated coefficient magnitudes across lags.

\begin{figure}[htbp]
\centering
\includegraphics[width=\textwidth]{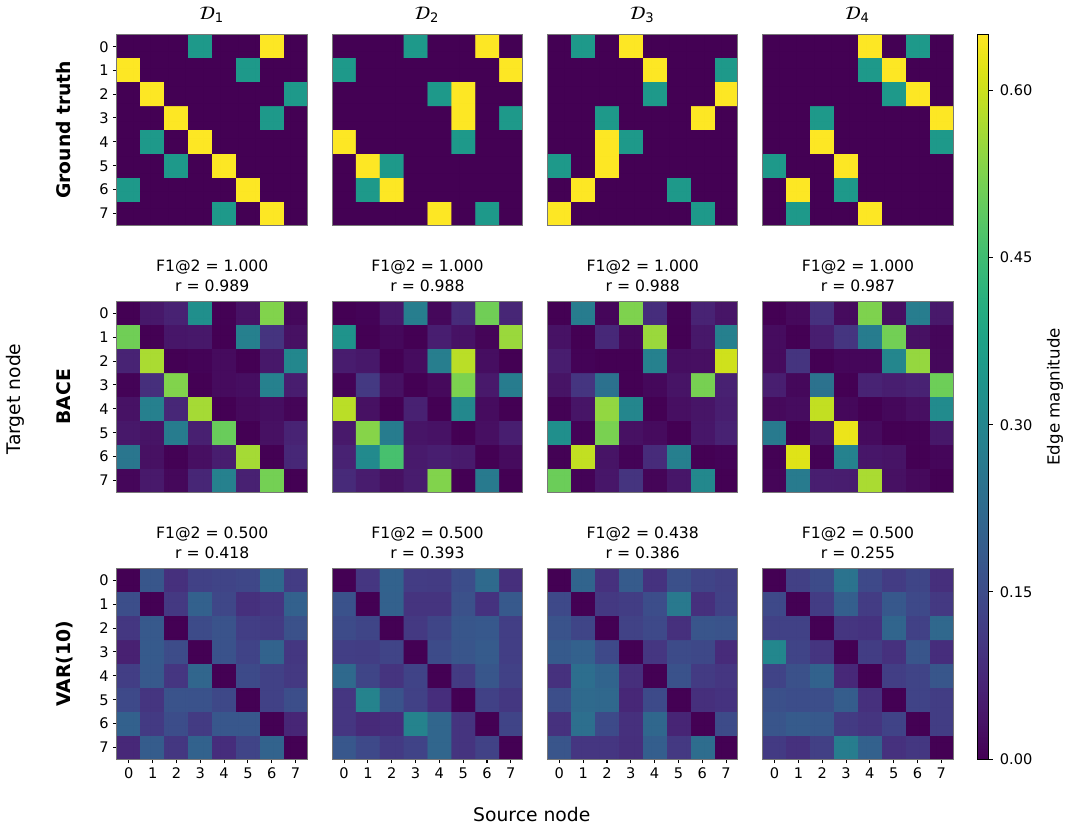}
\caption{\textbf{Graph recovery in an example nonlinear synthetic run.}
Columns correspond to four synthetic contexts.
Rows show absolute ground-truth weights, absolute BACE adjacency
weights, and VAR(10) edge scores.
Within each matrix, rows denote targets and columns denote sources;
all panels share a common color scale.
Annotations report \textsc{F1@}$k_{\text{row}}$ ($k=2$) and
Pearson correlation between edge magnitudes ($r$).}
\label{fig:nonlinear_recovery}
\end{figure}

\paragraph{Graph-recovery evaluation.}
Support recovery was assessed using \textsc{F1@}$k_{\text{row}}$ with $k=2$, matching the number of planted parents per target. Edge-magnitude agreement was assessed using Pearson correlation between absolute inferred and ground-truth edge weights. Metrics were averaged across the four contexts; the main text reports mean $\pm$ standard deviation across three independently generated datasets, and Fig.~\ref{fig:nonlinear_recovery} shows a representative run.

\subsection{Baseline Implementation Details}
\label{appendix:baseline_details}

\paragraph{Details of Comparison Models.}
Except for VAR/Granger, comparison models were trained and evaluated using the same participant-specific train/validation/test splits as BACE. Unless otherwise stated, models used the same input window length ($T_{\mathrm{in}}=100$), forecasting horizon ($T_{\mathrm{out}}=20$), and 80 retained channels. Hyperparameters were selected using validation performance, and the checkpoint with the best validation loss was used for test evaluation.
\paragraph{LSTM.} 
We used a two-layer LSTM forecaster with hidden dimension 128. The model takes past windows across all 80 channels and predicts the next 20 samples jointly for every channel. Training used mean squared error loss with Adam, learning rate $10^{-3}$, weight decay $10^{-4}$, and gradient clipping at 1.0. Models were trained with early stopping based on validation loss.
\paragraph{Context-aware LSTM.}
To control for access to behavioral context labels, we also trained a context-aware LSTM. This model uses the same LSTM architecture as the standard LSTM, but receives an additional learned embedding of the behavioral context label. This embedding is concatenated to the input representation before forecasting. The model therefore tests whether providing behavioral context alone improves prediction without learning explicit context-specific graphs.
\paragraph{NDT.} 
We implemented a Neural Data Transformer (NDT) baseline under a masked-forecasting setup. Although NDT is typically used to reconstruct randomly masked neural activity, we adapted it for forecasting by treating the final 20 samples of each window as the masked segment. The model uses a three-layer Transformer with attention dimension 128 over windows of length $100+20$. During training, the past 100 samples are provided and the final 20 samples are zero-masked; the model is optimized to reconstruct only the masked future segment. Training used Adam with learning rate $10^{-4}$, weight decay $10^{-5}$, and early stopping based on validation loss.

\paragraph{Graph WaveNet.} 
We implemented Graph WaveNet (GWNet) as a graph-based forecasting baseline for the cued reaching LFP datasets from the 3 participants. The architecture consists of 3 layers, each composed of 2 spatiotemporal blocks, for 6 blocks in total. Each block includes a gated dilated temporal convolution, a graph convolutional module for spatial mixing, and residual and skip connections. Temporal convolutions use kernel size 2, with 64 channels in the residual and dilated paths and 128 channels in the skip path. The skip outputs are aggregated and passed through a two-stage output head consisting of a $1\times1$ convolution with 64 channels followed by a readout convolution mapping to the 20-step forecasting horizon. Training was performed with Adam using an initial learning rate of $5\times10^{-4}$, decayed by 0.95 every 50 epochs, and no weight decay. Static graph supports were constructed from training data using correlation-based $k$-nearest neighbors, and an adaptive adjacency matrix was learned jointly during training.

\paragraph{DCRNN.} 
We implemented the Diffusion Convolutional Recurrent Neural Network (DCRNN) using a sequence-to-sequence architecture with diffusion-convolutional GRU (DCGRU) cells. Both the encoder and decoder consisted of two stacked DCGRU layers with hidden dimension 64. Each DCGRU layer used diffusion convolution with $K=2$ steps. The model was trained with Adam using learning rate $5\times10^{-4}$, learning-rate decay of 0.95 every 50 epochs, no weight decay, and inverse-sigmoid scheduled sampling with $\tau=3000$.

\paragraph{AMAG.}  
We implemented AMAG as a channel-level adaptive graph forecasting baseline. The 80 retained LFP channels were treated as 80 graph nodes, with one signal feature per node. Following the AMAG design, the model used three modules: temporal encoding, spatial interaction, and temporal readout. The spatial interaction module learned additive and multiplicative adjacency matrices and used a sample-dependent adaptor to modulate interactions between channel pairs. The adaptor was computed from a compact 5-step pooled summary of the past encoder states, while the temporal encoder and temporal readout operated on the full $T_{\mathrm{in}}=100$ input window and $T_{\mathrm{out}}=20$ forecasting horizon. Additive and multiplicative adjacency matrices were initialized from channel-wise correlations computed on the training data only. The temporal encoder and temporal readout used single-layer Transformer modules with hidden size 64 and 2 attention heads. Training used Adam with initial learning rate $5\times10^{-4}$, learning-rate decay of 0.95 every 50 epochs, weight decay $10^{-5}$, gradient clipping at 1.0, and early stopping based on validation loss.
\paragraph{Context-aware AMAG.}
We also implemented a context-aware AMAG variant to control for access to behavioral context labels. This model used the same channel-level AMAG architecture described above, with 80 retained LFP channels treated as graph nodes. The only modification was the addition of a learned embedding for the behavioral context label, which was added to the input representation before forecasting. All other components, including the temporal encoding, spatial interaction, temporal readout, additive and multiplicative adjacency matrices, and training procedure, were kept the same as in AMAG.
\subsection{Compute Resources and Computational Complexity}
\label{appendix:compute}

All neural forecasting models were profiled on a single workstation equipped with an AMD Ryzen Threadripper PRO~7955WX CPU with 16 cores, 128~GiB RAM, and one NVIDIA RTX~4500 Ada GPU with 24~GiB VRAM. Training and profiling were performed using PyTorch 2.2 with CUDA~12.2.

We compare the number of trainable parameters, \#P(M), and peak GPU memory, $M$ (GB), for neural forecasting models in Table~\ref{tab:complexity}. Parameter counts were computed from trainable model weights. Peak GPU memory was measured during training-style forward and backward passes using the same batch size and input/output window lengths used in the forecasting experiments.

\begin{table}[h]
\centering
\caption{Computation complexity for neural forecasting models. Parameter count and peak GPU memory are reported under the same training setup.}
\label{tab:complexity}
\vspace{1mm}
\footnotesize
\setlength{\tabcolsep}{3pt}
\renewcommand{\arraystretch}{0.88}
\begin{tabular}{lccccccc}
\toprule
 & LSTM & NDT & GWNet & DCRNN & AMAG & Ctx-AMAG & BACE \\
\midrule
\#P (M) & 0.45 & 0.63 & 0.48 & 0.30 & 0.253 & 0.291 & \textbf{0.13} \\
Peak memory (GB) & \textbf{0.15} & 0.84 & 5.54 & 5.22 & 17.26 & 17.26 & 0.20 \\
\bottomrule
\end{tabular}
\vspace{-2mm}
\end{table}

BACE has a small computational footprint, with 0.13M trainable parameters and 0.20 GB peak GPU memory, while also achieving the best forecasting results in the main benchmark table. This efficiency follows from its region-level design: BACE learns compact $N\times N$ context-specific graphs and maps the resulting regional representation back to the channel level for forecasting.

Among graph-based baselines, AMAG and context-aware AMAG have moderate parameter counts but require substantially more GPU memory in our implementation. GWNet and DCRNN also use larger graph-based operations than BACE. In contrast, BACE keeps the temporal encoder and forecasting head shared across behavioral contexts, and only the small adjacency matrices vary with context. These results indicate that BACE's gains arise from its architectural design rather than increased model capacity, combining efficient forecasting with interpretable graph outputs.

For BACE, the reported runs used 200 training epochs. A training-style epoch, measured using forward and backward passes for profiling, required approximately 27.15 seconds on the NVIDIA RTX~4500 Ada GPU. Validation and test forward-only evaluation required approximately 2.06 and 2.12 seconds, respectively.

\section{Additional Dataset and Analysis Details}
\label{appendix:experiments}

\subsection{Human LFP Dataset and Preprocessing}
\label{appendix:data_details}

\paragraph{Participants and experimental protocol.}
We analyzed three human intracranial local field potential (LFP) datasets collected from pediatric male participants undergoing deep brain stimulation (DBS) treatment as part of clinical care \citep{sanger2018case,kasiri2023correlated,Mousavitiming,kasiri2025optimized}. All recordings were collected under Institutional Review Board-approved protocols, with informed consent obtained from the participants' legal guardians.

All participants performed a cued reaching task. At the beginning of each trial, the participant held a home button. After a variable delay, a visual go cue prompted the participant to release the home button and reach forward to press a target button. After pressing the target, the participant returned the hand to the home button to prepare for the next trial. Trials with incomplete task execution or excessive artifacts were excluded before model fitting.

\paragraph{Recording electrodes and acquisition.}
Recordings were obtained from up to 10 temporary stereoelectroencephalography (sEEG) depth leads (AdTech MM16C), implanted into candidate DBS targets using standard stereotactic procedures \citep{sanger2018case}. Each lead included ten high-impedance micro-contacts (50~$\mu$m, 70--90 k$\Omega$). Signals were digitized at 24,414 Hz with a PZ5M 256-channel digitizer and RZ2 processor, and stored on an RS4 system (Tucker-Davis Technologies) \citep{kasiri2025local,soroushmojdehi2025frequency}. Electrode coverage differed across participants because implantation was clinically determined. Therefore, contacts were grouped into participant-specific anatomical nodes before model training, and BACE was trained separately for each participant.

Table~\ref{tab:participant_data_details} summarizes the node definitions and retained channels for each dataset.

\begin{table}[H]
\centering
\caption{Participant-level recording and preprocessing details. Internal clinical identifiers are not shown.}
\label{tab:participant_data_details}
\footnotesize
\setlength{\tabcolsep}{3pt}
\renewcommand{\arraystretch}{1.15}
\begin{tabularx}{\linewidth}{
@{}
>{\centering\arraybackslash}p{0.055\linewidth}
>{\centering\arraybackslash}p{0.085\linewidth}
>{\centering\arraybackslash}p{0.085\linewidth}
>{\centering\arraybackslash}p{0.055\linewidth}
>{\centering\arraybackslash}p{0.120\linewidth}
>{\raggedright\arraybackslash}X
@{}}
\toprule
Dataset &
Original channels &
Retained channels &
Nodes &
Trials/context after cleaning &
Anatomical nodes \\
\midrule
P1 & 80 & 80 & 8 & 326 &
GPi anterior-L/R, GPi posterior-L/R, VIM-L/R, STN-L/R \\

P2 & 100 & 80 & 10 & 329 &
GPi anterior-L/R, GPi posterior-L/R, VOa--VOp-L/R, VIM-L/R, CMCL-L/R \\

P3 & 100 & 80 & 10 & 398 &
CMP-L/R, GPi anterior-L/R, GPi posterior-L/R, VIM-L/R, VOa--VOp-L/R \\
\bottomrule
\end{tabularx}

\vspace{1mm}
\begin{flushleft}
\footnotesize
Abbreviations: GPi, globus pallidus internus; VIM, ventral intermediate nucleus; STN, subthalamic nucleus; VOa--VOp, ventral oral anterior/posterior thalamic nuclei; CMCL, centromedian--central lateral thalamic region; CMP, centromedian--parafascicular thalamic region.
\end{flushleft}
\end{table}

\paragraph{Preprocessing and cleaning.}
Signals were band-pass filtered from 1--50~Hz, common-average referenced, and downsampled to 1~kHz. Each dataset was cleaned before model fitting. For datasets with 100 original channels, two noisy channels were removed from each anatomical group, yielding eight retained contacts per node and 80 retained channels in total. Outlier trials were removed using robust amplitude-based criteria, and the final retained trials were balanced across behavioral contexts within each participant.

\paragraph{Trial segmentation.}
Each trial was segmented into four behavioral contexts of equal duration \citep{asadi2025graph}:
\begin{itemize}
    \item \textbf{Wait:} 400~ms preceding go cue onset.
    \item \textbf{React:} 0--400~ms following go cue onset, before movement initiation.
    \item \textbf{Reach:} 0--400~ms after forward reach onset, defined as the start of movement toward the target button.
    \item \textbf{Return:} 0--400~ms after return onset, defined as the start of movement from the target button back to the home button.
\end{itemize}
This segmentation produced event-aligned windows for comparing predictive connectivity across behavioral contexts.

\paragraph{Sliding windows, splitting, and normalization.}
Within each 400~ms behavioral context, data were divided into sliding forecasting examples with $T_{\mathrm{in}}=100$ input samples, $T_{\mathrm{out}}=20$ prediction samples, and stride 20. Train/validation/test splits were performed at the trial level rather than the window level to prevent leakage between overlapping windows from the same trial. Within each behavioral context, trials were shuffled using a fixed split seed and divided into 70\% training, 15\% validation, and 15\% test sets, and the same split was used for BACE, ablations, and baseline models. All trials for a given participant were recorded during a single session on one day. Normalization parameters were computed from the training trials only and then applied to validation and test trials.

\subsection{Graph and Context Ablation Analyses}
\label{appendix:ablation_details}

\paragraph{Implementation of the graph-free controls.}
In NoGraph and Ctx-NoGraph, the regional encoders process each
region's input independently, decoder attention operates over time
separately within each region, and each regional output is mapped
to that region's own channels. Both variants omit adjacency-based
message passing and autoregressive graph feedback; Ctx-NoGraph
additionally receives the behavioral-context label through the
decoder. This preserves context-conditioned regional forecasting
while removing the explicit graph-mediated routes for exchanging
information between regions.

\paragraph{Source-region perturbation analysis.}
We interpret $\mathbf{A}^{(k)}$ as describing how regions exchange information (regional effective connectivity) within the BACE model. We therefore checked whether this exchange occurs only through the pathways that use $\mathbf{A}^{(k)}$. If information could also pass between regions through other parts of the model, such as the encoders, temporal attention, or decoder, then $\mathbf{A}^{(k)}$ would capture only part of the model's inter-regional structure.

This concern leads to a direct, testable prediction. In Full BACE, perturbing the input of one region can change the forecasts of other regions because that perturbation can travel through the graph pathways. In contrast, if these are the only routes for regions to interact, then in a matched model with these pathways removed, the same perturbation should change only the perturbed region's own forecast and leave every other region unaffected. Ctx-NoGraph provides this control: it is a matched ablation that retains the regional encoders, context conditioning, attention mechanism, decoder, and forecasting objective, but removes the $\mathbf{A}^{(k)}$-mediated graph pathways---both graph mixing and graph feedback.

To test this prediction, we perturbed one source region at a time on held-out examples by replacing all channels and time samples of its normalized input history with zero, corresponding to the training mean, while leaving the behavioral-context label and all other regions' inputs unchanged. For each source--target pair, we measured the mean absolute change in the target region's forecast, averaged over held-out examples, forecast steps, and channels within the target region. Changes in the perturbed region's own forecast measure the local effect; changes in other regions' forecasts measure cross-region propagation.

The results matched the prediction (Table~\ref{tab:source_perturbation}). Local effects were similar in the two models, confirming that both responded to the perturbation to a similar degree. In Full BACE, the perturbation also changed the forecasts of other regions for all three participants. In Ctx-NoGraph, the effect stayed entirely within the perturbed region, with zero cross-region change to numerical precision. Removing the graph pathways therefore eliminated the observed cross-region effects while preserving a similar local response, supporting the conclusion that in the fitted models, regions exchange information only through the adjacency-mediated pathways.

Finally, we tested whether the size of each cross-region effect followed the strength of the corresponding learned edge. Within each participant and behavioral context, we correlated the off-diagonal forecast changes with the edge magnitudes $|A^{(k)}_{ij}|$, where $j$ is the perturbed source and $i$ the target. Correlations were consistently high (Pearson $r$ = 0.914--0.953 for P1, 0.903--0.957 for P2, and 0.928--0.947 for P3): perturbing a source region changed a target's forecast more when the learned edge between them was stronger. Together, these results support the conclusion that inter-regional interaction in BACE is both confined to the adjacency-mediated pathways and organized by the learned adjacency, consistent with interpreting $\mathbf{A}^{(k)}$ as the model's directed predictive connectivity between regions.

\begin{table}[h]
\centering
\caption{Source-region perturbation on held-out examples.
Values are mean absolute changes in normalized forecasts,
pooled across the four behavioral contexts. Local effects
refer to the perturbed region's own forecast; cross-region
effects refer to forecasts of other regions. Ctx-NoGraph
cross-region effects were zero to numerical precision.}
\label{tab:source_perturbation}
\vspace{1mm}
\small
\setlength{\tabcolsep}{7pt}
\renewcommand{\arraystretch}{1.05}
\begin{tabular}{lcccc}
\toprule
& \multicolumn{2}{c}{Full BACE}
& \multicolumn{2}{c}{Ctx-NoGraph} \\
\cmidrule(lr){2-3}
\cmidrule(lr){4-5}
Participant & Local & Cross-region & Local & Cross-region \\
\midrule
P1 & 0.580 & 0.017 & 0.584 & 0.000 \\
P2 & 0.559 & 0.011 & 0.556 & 0.000 \\
P3 & 0.561 & 0.020 & 0.579 & 0.000 \\
\bottomrule
\end{tabular}
\end{table}

\subsection{Statistical analysis of learned connectivity}
\label{appendix:stats}

\paragraph{Split--half reproducibility.}
For each participant and behavioral context $k$, we performed $K=50$ random half-splits of trials. For each split, BACE graphs were estimated separately on each half, and reproducibility was measured as the Pearson correlation between the vectorized absolute adjacency matrices $|\mathbf{A}^{(k)}|$ from the two halves. We then applied the Spearman--Brown correction,
\[
r_{\mathrm{SB}}=\frac{2r}{1+r},
\]
to estimate full-data repeatability.

Mean split-half correlations were:
P1: Wait $r=0.91$, React $r=0.90$, Reach $r=0.88$, Return $r=0.85$;
P2: Wait $r=0.90$, React $r=0.88$, Reach $r=0.91$, Return $r=0.83$;
P3: Wait $r=0.90$, React $r=0.91$, Reach $r=0.89$, Return $r=0.84$.

The corresponding Spearman--Brown corrected repeatability values were:
P1: 0.95, 0.95, 0.94, and 0.92;
P2: 0.95, 0.94, 0.95, and 0.91;
P3: 0.95, 0.95, 0.94, and 0.91.
These values indicate that BACE's context-specific graphs are reproducible across different subsets of trials.

\paragraph{Bootstrap resampling.} 
To quantify uncertainty in individual edges, we performed $B=1000$ nonparametric bootstrap resamples within each behavioral context. For each bootstrap sample, we re-estimated the graph parameters while holding the trained encoder, graph projector, and forecasting head fixed. This procedure measures how stable the learned adjacency estimates are under trial resampling. For each directed edge $(j\!\to\! i)$, we computed percentile 95\% confidence intervals for the absolute edge strength $|A^{(k)}_{ij}|$.

\paragraph{Edgewise precision.} 
We summarized bootstrap uncertainty using the width of the 95\% confidence interval for each edge. Because very small median edge values can make relative uncertainty unstable, we computed relative half-width only for edges whose median magnitude exceeded a threshold $\tau$, set to the 66th percentile of edge magnitudes:
\[
\mathrm{rHW} = \frac{\mathrm{hi}-\mathrm{lo}}{2\,|\mathrm{median}|}.
\]
This analysis provides a scale-normalized measure of uncertainty for non-negligible edges and complements the across-context tests reported below.

\paragraph{Across-context edge differences.}
To test whether learned edge strengths differed across behavioral contexts, we compared each pair of context-specific adjacency matrices within each participant. For each ordered edge $(j\!\to\! i)$ and each context pair $(k_1,k_2)$, we computed the difference in absolute predictive edge strength,
\[
\Delta_{ij}^{(k_1,k_2)}
=
\left|A_{ij}^{(k_1)}\right|
-
\left|A_{ij}^{(k_2)}\right|.
\]
Uncertainty was estimated with nonparametric bootstrap resampling over trials within each context. For each bootstrap sample, BACE adjacency matrices were re-estimated and the edgewise difference $\Delta_{ij}^{(k_1,k_2)}$ was recomputed. Edges were considered context-dependent if the bootstrap confidence interval excluded zero and the corresponding two-sided test survived Benjamini--Hochberg FDR correction at $q=0.05$. Correction was applied within each participant and context-pair comparison across all directed edges. 

Fig.~\ref{fig:group_results}A shows an example participant, displaying only edges that survived this procedure.

\section*{Ethics statement}
Human neural data were collected under IRB-approved protocols at participating medical centers. Informed consent was obtained from participants or legal guardians for minors. All data were de-identified before analysis, stored on secure access-controlled servers, and used solely for research in compliance with HIPAA and institutional privacy policies. The study did not involve clinical decision-making, non-standard interventions, or recruitment beyond patients already evaluated for DBS, and no re-identification was attempted. Raw recordings cannot be publicly shared due to ethical and regulatory restrictions, but de-identified datasets may be provided upon reasonable request under IRB-approved data use agreements. The authors confirm adherence to the NeurIPS Code of Ethics and declare no conflicts of interest or external sponsorship that could have influenced the research.
%%%%%%%%%%%%%%%%%%%%%%%%%%%%%%%%%%%%%%%%%%%%%%%%%%%%%%%%%%%%

\newpage
\section*{NeurIPS Paper Checklist}

\begin{enumerate}

\item {\bf Claims}
    \item[] Question: Do the main claims made in the abstract and introduction accurately reflect the paper's contributions and scope?
    \item[] Answer: \answerYes{} % Replace by \answerYes{}, \answerNo{}, or \answerNA{}.
    \item[] Justification: The abstract and Introduction state that BACE learns behavior-conditioned effective connectivity among anatomically defined brain nodes and is evaluated on three synthetic datasets and three human intracranial datasets. The claims are supported by the experimental results and are stated within the scope of the evaluated datasets and tasks.

\item {\bf Limitations}
    \item[] Question: Does the paper discuss the limitations of the work performed by the authors?
    \item[] Answer: \answerYes{} % Replace by \answerYes{}, \answerNo{}, or \answerNA{}.
    \item[] Justification: The Discussion states that BACE has been evaluated on three human intracranial datasets, but that the cohort remains small and clinically heterogeneous. It also notes that the current model requires experimentally defined behavioral context labels, which limits its use to labeled datasets. Future work should test BACE on larger cohorts and other behavioral contexts. 

\item {\bf Theory assumptions and proofs}
    \item[] Question: For each theoretical result, does the paper provide the full set of assumptions and a complete (and correct) proof?
    \item[] Answer: \answerNA{} % Replace by \answerYes{}, \answerNo{}, or \answerNA{}.
    \item[] Justification: The paper does not present formal theoretical results or proofs. The mathematical expressions define the model architecture, forecasting objective, and evaluation setup.

    \item {\bf Experimental result reproducibility}
    \item[] Question: Does the paper fully disclose all the information needed to reproduce the main experimental results of the paper to the extent that it affects the main claims and/or conclusions of the paper (regardless of whether the code and data are provided or not)?
    \item[] Answer: \answerYes{} % Replace by \answerYes{}, \answerNo{}, or \answerNA{}.
    \item[] Justification: The paper describes the datasets, preprocessing pipeline, model architecture, training procedures, hyperparameters, evaluation metrics, baselines, and compute resources in the main text and appendix. The BACE project repository is available at \url{https://github.com/mehrnazasadi95/BACE}.

\item {\bf Open access to data and code}
    \item[] Question: Does the paper provide open access to the data and code, with sufficient instructions to faithfully reproduce the main experimental results, as described in supplemental material?
    \item[] Answer: \answerYes{} % Replace by \answerYes{}, \answerNo{}, or \answerNA{}.
    \item[] Justification: Because this study uses human intracranial recordings, the raw data cannot be publicly released due to ethical restrictions and institutional approvals. De-identified data may be shared upon reasonable request under IRB-approved data use agreements. Code for preprocessing, training, evaluation, and analysis, along with documentation, is  available at \url{https://github.com/mehrnazasadi95/BACE.}

\item {\bf Experimental setting/details}
    \item[] Question: Does the paper specify all the training and test details (e.g., data splits, hyperparameters, how they were chosen, type of optimizer) necessary to understand the results?
    \item[] Answer: \answerYes{} % Replace by \answerYes{}, \answerNo{}, or \answerNA{}.
    \item[] Justification: The paper specifies the trial-level train/validation/test splits, sliding-window construction, input and prediction horizons, evaluation metrics, optimizer, and checkpoint selection in the main text. The appendix provides additional details on preprocessing, normalization, BACE hyperparameters, graph initialization, and baseline implementations.

\item {\bf Experiment statistical significance}
    \item[] Question: Does the paper report error bars suitably and correctly defined or other appropriate information about the statistical significance of the experiments?
    \item[] Answer: \answerYes{} % Replace by \answerYes{}, \answerNo{}, or \answerNA{}.
    \item[] Justification: The paper reports variability measures and statistical analyses for the main experimental results, including mean and standard deviation across runs for forecasting and ablation experiments, split-half reproducibility for learned connectivity, bootstrap confidence intervals for edge reliability, and corrected p-values using Benjamini--Hochberg false discovery rate (FDR) correction. Additional participant-level reliability and statistical results are reported in the appendix.

\item {\bf Experiments compute resources}
    \item[] Question: For each experiment, does the paper provide sufficient information on the computer resources (type of compute workers, memory, time of execution) needed to reproduce the experiments?
    \item[] Answer: \answerYes{} % Replace by \answerYes{}, \answerNo{}, or \answerNA{}.
    \item[] Justification: The paper reports compute information for the main neural forecasting experiments, including CPU/GPU specifications, system and GPU memory, software environment, trainable parameter counts, peak GPU memory, and approximate execution time for BACE training, validation, and test evaluation. These details are summarized in the computation complexity table and compute-resources appendix.
    
\item {\bf Code of ethics}
    \item[] Question: Does the research conducted in the paper conform, in every respect, with the NeurIPS Code of Ethics \url{https://neurips.cc/public/EthicsGuidelines}?
    \item[] Answer: \answerYes{} % Replace by \answerYes{}, \answerNo{}, or \answerNA{}.
    \item[] Justification: The study was conducted under IRB-approved protocols, with informed consent obtained from participants or their legal guardians. Human neural data were de-identified before analysis, stored on secure access-controlled servers, and used only for research. 

\item {\bf Broader impacts}
    \item[] Question: Does the paper discuss both potential positive societal impacts and negative societal impacts of the work performed?
    \item[] Answer: \answerYes{} % Replace by \answerYes{}, \answerNo{}, or \answerNA{}.
    \item[] Justification: The paper discusses the broader implications of BACE as a general framework for learning interpretable, behavior-conditioned connectivity from neural time series. Potential positive impacts include supporting neuroscience research, enabling hypothesis generation about brain-region interactions during behavior, and providing a model that can be adapted to other multi-region neural recordings beyond the datasets studied here. The paper also discusses ethical considerations around human intracranial data, including privacy, controlled data sharing and cohort heterogeneity.
    
\item {\bf Safeguards}
    \item[] Question: Does the paper describe safeguards that have been put in place for responsible release of data or models that have a high risk for misuse (e.g., pre-trained language models, image generators, or scraped datasets)?
    \item[] Answer: \answerYes{} % Replace by \answerYes{}, \answerNo{}, or \answerNA{}.
    \item[] Justification: The paper does not release a high-risk pretrained model, generative model, or scraped dataset. The human intracranial recordings are not publicly released; access to de-identified data is controlled through reasonable request and execution of a data use agreement consistent with IRB approvals.

\item {\bf Licenses for existing assets}
    \item[] Question: Are the creators or original owners of assets (e.g., code, data, models), used in the paper, properly credited and are the license and terms of use explicitly mentioned and properly respected?
    \item[] Answer: \answerYes{} % Replace by \answerYes{}, \answerNo{}, or \answerNA{}.
    \item[] Justification: Existing assets used in the paper, including open-source machine learning frameworks and scientific computing libraries such as PyTorch and standard Python analysis packages, are used in accordance with their respective licenses. The paper also cites the existing methods and models used as baselines. The human intracranial recordings were collected under IRB-approved clinical research protocols and are not publicly redistributed due to ethical and privacy restrictions.

\item {\bf New assets}
    \item[] Question: Are new assets introduced in the paper well documented and is the documentation provided alongside the assets?
    \item[] Answer: \answerYes{} % Replace by \answerYes{}, \answerNo{}, or \answerNA{}.
    \item[] Justification: The paper introduces and releases a code repository, including preprocessing, training, and evaluation code, along with documentation, environment specifications, and synthetic data examples. The human intracranial recordings are not released as a public asset because of ethical and privacy restrictions, but de-identified data may be shared upon reasonable request under an IRB-consistent data use agreement.

\item {\bf Crowdsourcing and research with human subjects}
    \item[] Question: For crowdsourcing experiments and research with human subjects, does the paper include the full text of instructions given to participants and screenshots, if applicable, as well as details about compensation (if any)? 
    \item[] Answer: \answerYes{} % Replace by \answerYes{}, \answerNo{}, or \answerNA{}.
    \item[] Justification: The paper analyzes previously collected human intracranial recordings obtained under IRB-approved protocols with informed consent. The authors did not conduct new participant recruitment, crowdsourcing, surveys, and compensation. The task protocol and trial structure are described in the dataset and preprocessing appendix.

\item {\bf Institutional review board (IRB) approvals or equivalent for research with human subjects}
    \item[] Question: Does the paper describe potential risks incurred by study participants, whether such risks were disclosed to the subjects, and whether Institutional Review Board (IRB) approvals (or an equivalent approval/review based on the requirements of your country or institution) were obtained?
    \item[] Answer: \answerYes{} % Replace by \answerYes{}, \answerNo{}, or \answerNA{}.
    \item[] Justification: As described in the Ethics statement, the human intracranial recordings analyzed in this study were collected under IRB-approved protocols with informed consent from participants or their legal guardians. Potential risks associated with the original clinical recordings were governed by the approved clinical research protocols. All data used in this work were de-identified before analysis, stored securely, and used only for research.

\item {\bf Declaration of LLM usage}
    \item[] Question: Does the paper describe the usage of LLMs if it is an important, original, or non-standard component of the core methods in this research? Note that if the LLM is used only for writing, editing, or formatting purposes and does \emph{not} impact the core methodology, scientific rigor, or originality of the research, declaration is not required.
    %this research? 
    \item[] Answer: \answerNA{} % Replace by \answerYes{}, \answerNo{}, or \answerNA{}.
    \item[] Justification: LLMs were not used in developing the BACE model, running experiments, analyzing data, implementing baselines, or producing the reported results. Any AI-assisted tools were limited to language polishing, editing, and formatting, and did not affect the methodology, experimental design, or scientific conclusions.

\end{enumerate}

\end{document}